# Model-driven Development of Complex Software: A Research Roadmap

Robert France, Bernhard Rumpe

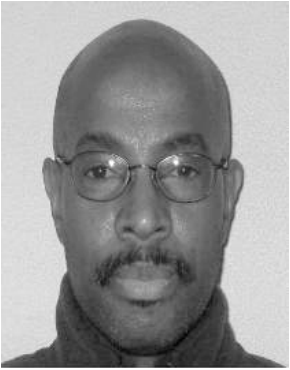

Robert France is a Professor in the Department of Computer Science at Colorado State University. His research focuses on the problems associated with the development of complex software systems. He is involved in research on rigorous software modeling, on providing rigorous support for using design patterns, and on separating concerns using aspect-oriented modeling techniques. He was involved in the Revision Task Forces for UML 1.3 and UML 1.4. He is currently a Co-Editor-In-Chief for the Springer international journal on Software and System Modeling, a Software Area Editor for IEEE Computer and an Associate Editor for the Journal on Software Testing, Verification and Reliability.

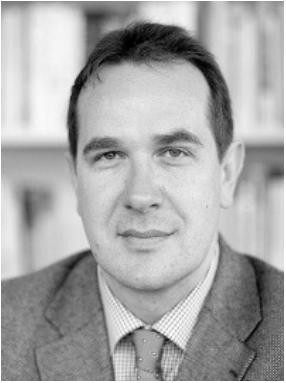

Bernhard Rumpe is chair of the Institute for Software Systems Engineering at the Braunschweig University of Technology, Germany. His main interests are software development methods and techniques that benefit form both rigorous and practical approaches. This includes the impact of new technologies such as model-engineering based on UML-like notations and domain specific languages and evolutionary, test-based methods, software architecture as well as the methodical and technical implications of their use in industry. He has furthermore contributed to the communities of formal methods and UML. He is author and editor of eight books and Co-Editor-in-Chief of the Springer International Journal on Software and Systems Modeling (www.sosym.org).

# Model-driven Development of Complex Software: A Research Roadmap

Robert France
Department of Computer Science
Colorado State University
Fort Collins, CO 80523
france@cs.colostate.edu

Bernhard Rumpe
Software Systems Engineering Institute
Faculty for Mathematics and Computer Science
Braunschweig University of Technology
Braunschweig, Germany
http://www.sse.cs.tu-bs.de

## Abstract

*The term Model-Driven Engineering (MDE) is typically used to describe software development approaches in which abstract models of software systems are created and systematically transformed to concrete implementations. In this paper we give an overview of current research in MDE and discuss some of the major challenges that must be tackled in order to realize the MDE vision of software development. We argue that full realizations of the MDE vision may not be possible in the near to medium-term primarily because of the wicked problems involved. On the other hand, attempting to realize the vision will provide insights that can be used to significantly reduce the gap between evolving software complexity and the technologies used to manage complexity.*

## 1. Introduction

Advances in hardware and network technologies have paved the way for the development of increasingly pervasive software-based systems of systems that collaborate to provide essential services to society. Software in these systems is often required to (1) operate in distributed and embedded computing environments consisting of diverse devices (personal computers, specialized sensors and actuators), (2) communicate using a variety of interaction paradigms (e.g., SOAP messaging, media streaming), (3) dynamically adapt to changes in operating environments, and (4) behave in a dependable manner [26, 62]. Despite significant advances in programming languages and supporting integrated development environments (IDEs), developing these complex software systems using current code-centric technologies requires herculean effort.

A significant factor behind the difficulty of developing complex software is the wide conceptual gap between the problem and the implementation domains of discourse. Bridging the gap using approaches that require extensive handcrafting of implementations gives rise to accidental complexities that make the development of complex software difficult and costly. To an extent, handcrafting complex software systems can be likened to building pyramids in ancient Egypt. We marvel at these software implementations in much the same way that archaeologists marvel at the pyramids: The wonder is mostly based on an appreciation of the effort required to tackle the significant accidental complexities arising from the use of inadequate technologies.

The growing complexity of software is the motivation behind work on industrializing software development. In particular, current research in the area of *model driven engineering* (MDE) is primarily concerned with reducing the gap between problem and software implementation domains through the use of technologies that support systematic transformation of problem-level abstractions to software implementations. The complexity of bridging the gap is tackled through the use of models that describe complex systems at multiple levels of abstraction and from a variety of perspectives, and through automated support for transforming and analyzing models. In the MDE vision of software development, models are the primary artifacts of development and developers rely on computer-based technologies to transform models to running systems.

Current work on MDE technologies tends to focus on producing implementation and deployment artifacts from detailed design models. These technologies use models to generate significant parts of (1) programs written in languages such as Java, C++, and C♯ (e.g., see Compuware's OptimalJ, IBM's Rational XDE package, and Microsoft's Visual Studio), and (2) integration and deployment artifacts such as XML-based configuration files and data bridges used for integrating disparate systems (e.g., see [25]).

Attempts at building complex software systems that dynamically adapt to changes in their operating environments has led some researchers to consider the use of models dur-ing runtime to monitor and manage the executing software. Early work in this emerging MDE area was presented at a MODELS 2006 Workshop on runtime models [8].

We envisage that MDE research on runtime models will pave the way for the development of environments in which change agents (e.g., software maintainers, software-based agents) use runtime models to modify executing software in a controlled manner. The models act as interfaces that change agents can use to adapt, repair, extend, or retrofit software during its execution. In our broad vision of MDE, models are not only the primary artifacts of development, they are also the primary means by which developers and other systems understand, interact with, configure and modify the runtime behavior of software.

A major goal of MDE research is to produce technologies that shield software developers from the complexities of the underlying implementation platform. An implementation platform may consist of networks of computers, middleware, and libraries of utility functions (e.g., libraries of persistence, graphical user interface, and mathematical routines). In the case of MDE research on runtime models, the goal is to produce technologies that hide the complexities of runtime phenomena from agents responsible for managing the runtime environment, and for adapting and evolving the software during runtime.

Realizing the MDE vision requires tackling a wide range of very difficult interrelated social and technical problems that has been the focus of software engineering research over the last three decades. For this reason, we consider the problem of developing MDE technologies that automate significant portions of the software lifecycle to be a wicked problem. A wicked problem has multiple dimensions that are related in complex ways and thus cannot be solved by cobbling solutions to the different problem dimensions (see definition of "wicked problem" in Wikipedia). Solutions to wicked problems are expensive to develop and are invariably associated with other problems, but the development of the solutions can deepen understanding of the problems.

In this paper we discuss some of the major technical problems and challenges associated with realizing the broad MDE vision we outline above. We also mention the social challenges related to identifying and leveraging high-quality modeling experience in MDE research.

In Section 2 we describe software development as a modeling activity and present the research questions that should drive MDE research. In Section 3 we discuss the factors that contribute to the difficulty of bridging the gap between the problem and implementation domains, present classes of challenges and problems discussed in this paper, and discuss the relationship between MDE and other areas of software engineering research. In Section 4 we provide background on some of the major MDE initiatives. In Section 5, Section 6, and Section 7, we discuss MDE research challenges in the areas of modeling languages, separation of concerns, and model manipulation and management, respectively. Section 7 also includes a discussion on opportunities for using models during runtime. We conclude in Section 8 by outlining an idealistic vision of an MDE environment.

## 2. The Value of Modeling

In this paper, a model is an abstraction of some aspect of a system. The system described by a model may or may not exist at the time the model is created. Models are created to serve particular purposes, for example, to present a human understandable description of some aspect of a system or to present information in a form that can be mechanically analyzed (e.g., see [54, 30]).

It may seem that work on MDE centers on the development and use of the popular modeling language, UML (the Unified Modeling Language) [59]. The UML standardization effort has played a vital role in bringing together a community that focuses on the problem of raising the level of abstraction at which software is developed, but research around other modeling languages is contributing valuable MDE concepts, techniques, tools and experience. In this paper, MDE encompasses all research pertaining to the use of software models.

Non-UML modeling approaches that are included in our use of the MDE term include specifying systems using formal specification languages such as Alloy [32] and B [2], modeling and analyzing control system software using the math-based, high-level programming language Matlab/Simulink/Stateflow (e.g., see [34]), analyzing performance, load, safety, liveness, reliability, and other system properties using specialized modeling techniques (e.g., see [40]), and building models to analyze software risks (e.g., see [20, 27, 50]).

Source code can be considered to be a model of how a system will behave when executed. While we may draw inspiration from work on the development of programming languages and compilers, this paper is primarily concerned with the development and use of models other than source code. Specifically, we focus attention on the following two broad classes of models:

- *Development models*: These are models of software at levels of abstraction above the code level. Examples of development models are requirements, architectural, implementation and deployment models. MDE research has tended to focus on the creation and use of these models.

- *Runtime models*: These models present views of some aspect of an executing system and are thus abstractions of runtime phenomena. A growing number of MDE researchers have started to explore how models can be used to support dynamic adaptation of software-based systems.

As MDE research matures, the above classification may become dynamic, that is, development models may be used as runtime models and runtime models may be used to evolve software systems, thus acting as development models.

There is a perception that development models are primarily documentation artifacts and thus their creation and use are peripheral to software development. This narrow perspective has led to recurring and seemingly futile debates on the practical value of modeling (i.e., the value of documentation) in software development. MDE advocates point out that models can be beneficially used for more than just documentation during development. For example, Bran Selic, an IBM Distinguished Engineer, points out an important property of software that MDE seeks to exploit[1]: "*Software has the rare property that it allows us to directly evolve models into full-fledged implementations without changing the engineering medium, tools, or methods.*" Selic and others argue that modeling technologies leveraging this property can significantly reduce the accidental complexities associated with handcrafting complex software [60].

The formal methods community attempted to leverage this property in work on transformation-based software development in which declarative specifications are systematically transformed to programs (e.g., see [7]). One of the valuable insights gained from these attempts is that automation of significant aspects of the transformation of a high-level specification to an implementation requires encoding domain-specific knowledge in the transformation tools. Challenges associated with developing and using domain-specific technologies will be discussed in Section 5.

The process of analyzing a problem, conceiving a solution, and expressing a solution in a high-level programming language can be viewed as an implicit form of modeling and thus one can argue that software development is essentially a model-based problem solving activity. The mental models of the system held by developers while creating programs may be shared with others using informal "whiteboard" sketches or more formally as statements (including diagrams) in a modeling language. These mental models evolve as a result of discussions with other developers, changes in requirements, and errors identified in code tests, and they guide the development of handcrafted code. Writing source code is a modeling activity because the developer is modeling a solution using the abstractions provided by a programming language.

Given that the technical aspects of software development are primarily concerned with creating and evolving models,

---

[1] taken from a presentation at the ENSIETA Summer School on Model-Driven Engineering of Embedded Systems, September 2004

questions about whether we should or should not use models seem superfluous. A more pertinent question is "Can modeling techniques be more effectively leveraged during software development?". From this perspective, the research question that should motivate MDE research on creation and use of development models is the following:

> How can modeling techniques be used to tame the complexity of bridging the gap between the problem domain and the software implementation domain?

Henceforth, we will refer to this gap as the *problem-implementation gap*.

We propose that MDE research on runtime models focus on the following research questions:

- How can models be cost-effectively used to manage executing software? Management can involve monitoring software behavior and the operating context, and adapting software so that it can continue to provide services when changes are detected in operating conditions.

- How can models be used to effect changes to running systems in a controlled manner? Research in this respect will focus on how models can be used as interfaces between running systems and change agents, where a change agent can be a human developer or a software agent.

There is currently very little work on the runtime modeling questions and thus there is very little research experience that can be used to bound possible solutions. We discuss some of the challenges we envisage in Section 7.3, but this paper focuses on development models primarily because current MDE research provides significant insights into associated challenges and problems.

## 3. MDE Research Concerns

MDE research on development models focuses on developing techniques, methods, processes and supporting tools that effectively narrow the problem-implementation gap. Exploring the nature of the problem-implementation gap can yield insights into the problems and challenges that MDE researchers face.

### 3.1. Bridging the Gap

A problem-implementation gap exists when a developer implements software solutions to problems using abstractions that are at a lower level than those used to express the problem. In the case of complex problems, bridging the gap using methods that rely almost exclusively on human effort will introduce significant accidental complexities [60].

The introduction of technologies that effectively raise the implementation abstraction level can significantly improve productivity and quality with respect to the types of software targeted by the technologies. The introduction and successful use of the technologies will inevitably open the door to new software opportunities that are acted upon. The result is a new generation of more complex software systems and associated software development concerns. For example, the introduction of middleware technologies, coupled with improvements in network and mobile technologies, has made it possible to consider the development of more complex distributed systems involving fixed and mobile elements.

The growing complexity of newer generations of software systems can eventually overwhelm the available implementation abstractions, resulting in a widening of the problem-implementation gap. The widening of the gap leads to dependence on experts who have built up an arsenal of mentally-held development patterns (a.k.a "experience") to cope with growing complexity.

Growing software complexity will eventually overwhelm the mentally-held experience and the need for technologies that leverage explicit forms of experience (e.g., domain-specific design patterns) to further raise the level of abstraction at which software is developed will become painfully apparent. The development of such technologies will result in work on even more complex software systems, thus triggering another cycle of work on narrowing the problem-implementation gap.

The preceding discussion indicates that research on narrowing the problem-implementation gap tends to progress through a series of crises-driven cycles. Each cycle results in a significant change in the level of abstraction at which software is developed, which then triggers attempts at building even more complex software. This characterization acknowledges that software development concerns and challenges evolve with each new generation of software systems, that is, the nature of the so-called "software crisis" evolves.

To cope with the ever-present problem of growing software complexity MDE researchers need to develop technologies that developers can use to generate domain-specific software development environments. These environments should consist of languages and tools that are tailored to the target classes of applications. Developing such technologies requires codifying knowledge that reflects a deep understanding of the common and variable aspects of the gap bridging process. Such an understanding can be gained only through costly experimentation and systematic accumulation and examination of experience. Developing such technologies is thus a wicked problem.

While it may not be possible to fully achieve the MDE vision, close approximations can significantly improve our ability to manage the problem-implementation gap. We see no alternative to developing close approximations other than through progressive development of technologies, where each new generation of technologies focuses on solving the problems and minimizing the accidental complexities arising from use of older generations of technologies.

The importance of industrial participation on MDE research should not be underestimated. Industrial feedback on techniques and technologies developed within academia is needed to gain a deeper understanding of development problems.

### 3.2. A Classification of MDE Challenges

The major challenges that researchers face when attempting to realize the MDE vision can be grouped into the following categories:

- Modeling language challenges: These challenges arise from concerns associated with providing support for creating and using problem-level abstractions in modeling languages, and for rigorously analyzing models.

- Separation of concerns challenges: These challenges arise from problems associated with modeling systems using multiple, overlapping viewpoints that utilize possibly heterogeneous languages.

- Model manipulation and management challenges: These challenges arise from problems associated with (1) defining, analyzing, and using model transformations, (2) maintaining traceability links among model elements to support model evolution and roundtrip engineering, (3) maintaining consistency among viewpoints, (4) tracking versions, and (5) using models during runtime.

Section 5 to Section 7 present some of the major challenges in these categories.

### 3.3. Relationship with Software Engineering Research

Realizing the MDE vision of software development will require evolving and integrating research results from different software engineering areas. There are obvious connections with work on requirements, architecture and detailed design modeling, including work on viewpoint conflict analysis and on feature interaction analysis. Research in these areas have produced modeling concepts, languages, and techniques that address specific concerns in the areas.

MDE research should leverage and integrate the best results from these areas and build synergistic research links with the communities. For example, the challenges faced by researchers in the software architecture area (see [58]) are closely related to MDE challenges and there have been beneficial interactions across the two communities.

Work on formal specification techniques (FSTs) is particularly relevant to MDE. Modeling languages must have formally defined semantics if they are to be used to create analyzable models. Work on developing formal analysis techniques for models utilizes and builds on work in the formal specification research area. While it is currently the case that popular modeling languages have poorly defined semantics, there is a growing realization that MDE requires semantic-based manipulation of models and thus appropriate aspects of modeling languages must be formalized.

It may seem that MDE research can be subsumed by FST research. A closer examination of research results and goals in these areas suggests that this is not the case. The FSTs that have been developed thus far use languages that allow developers to describe systems from a very small number of viewpoints. For example, Z [53] describes systems from data and operation viewpoints, model checking techniques (e.g., see [49]) are applicable to models created using a state transition viewpoint, and petri nets [48] can be used to describe systems from a control flow viewpoint. It is well known that the more expressive a modeling language is, the more intractable the problem of developing mechanical semantic analysis techniques becomes. It should not be surprising then that FSTs restrict their viewpoints.

In MDE, a model of a complex system consists of many views created using a wide variety of viewpoints. Furthermore, FSTs focus on describing functionality, while MDE approaches aim to provide support for modeling structural and functional aspects as well as system attributes (sometimes referred to as "non-functional" aspects).

The differences in research scopes indicate that MDE provides a context for FST research. There is often a need to formally analyze a subset of the views in an MDE model. Members of the FST and the MDE communities need to collaborate to produce formal techniques that can be integrated with rich modeling languages.

The following gives some of the other major software engineering research areas that influence MDE work:

- *Systematic reuse of development experience:* Leveraging explicit forms of development experience to industrialize software development has been the focus of research in the systematic reuse community for over two decades (e.g., see [4, 6, 21, 29, 33]). The term "software factory" was used in the systematic reuse community to refer to development environments that effectively leveraged reusable assets to improve productivity and quality [14]. The term is now being used by Microsoft as a label for its MDE initiative (see Section 4). Research on design patterns, domain-specific languages, and product-line architectures are particularly relevant to work on MDE (e.g., see [38]).

- *Systematic software testing:* Work on systematic testing of programs is being leveraged in work on dynamically analyzing models. For example, there is work on defining test criteria for UML models that are UML-specific variations of coverage criteria used at the code level [3], and tools that support systematic testing of models [18]. There is also work on generating code level tests from models that builds upon work in the specification-based code testing area (e.g., see [10, 43]).

- *Compilation technologies:* Work on optimizing compilers may be leveraged by MDE researchers working on providing support for generating lean and highly optimized code. Work on incremental compilation may also be leveraged in research on incremental code generation.

It can be argued that MDE is concerned with providing automated support for software engineering, and thus falls in the realm of computer-aided software engineering (CASE) research. MDE can and should be viewed as an evolution of early CASE work. MDE researchers are (knowingly or unknowingly) building on the experience and work of early CASE researchers. Unlike early CASE research, which focused primarily on the use of models for documenting systems (e.g. see [17, 19, 24]), MDE research is concerned with broadening the role of models so that they become the primary artifacts of software development. This broadening of the scope is reflected in the range of software engineering research areas that currently influence MDE research.

The need to deal with the complexity of developing and operating adaptive software provides another opportunity for the use of MDE techniques. In this paper, MDE encompasses work that seeks to develop a new generation of CASE environments that address the entire life-cycle of software, from conceptualization to retirement. It is concerned not only with the use of models for engineering complex software, but also with the use of models during the execution of software.

The term "model driven" may be considered by some to be redundant in MDE given that engineering of software invariably involves modeling. While this may be true, it is currently the case that software developers seldom create and effectively utilize models other than code. The term "model driven" in MDE is used to emphasize a shift away from code level abstractions. Only when modeling at various levels of abstraction is widely viewed as an essential part of engineering software should the "model driven" term be dropped. The availability of good modeling tools can help in this respect.

## 4. Major Model Driven Engineering Initiatives

In this section we present an overview of some major MDE initiatives that are currently shaping the research landscape and discuss the relationship between MDE and other software engineering research areas.

### 4.1. Model Driven Architecture

The OMG, in its role as an industry-driven organization that develops and maintains standards for developing complex distributed software systems, launched the Model Driven Architecture (MDA) as a framework of MDE standards in 2001 [60, 51]. The OMG envisages MDA technologies that will provide the means to more easily integrate new implementation infrastructures into existing designs, generate significant portions of application-specific code, configuration files, data integration bridges and other implementation infrastructure artifacts from models, more easily synchronize the evolution of models and their implementations as the software evolves, and rigorously simulate and test models.

MDA advocates modeling systems from three viewpoints: computation independent, platform independent, and platform specific viewpoints. The computation independent viewpoint focuses on the environment in which the system of interest will operate in and on the required features of the systems. Modeling a system from this viewpoint results in a computation independent model (CIM). The platform independent viewpoint focuses on the aspects of system features that are not likely to change from one platform to another. A platform independent model (PIM) is used to present this viewpoint. The OMG defines a platform as "a set of subsystems and technologies that provide a coherent set of functionality through interfaces and specified usage patterns". Examples of platforms are technology-specific component frameworks such as CORBA and J2EE, and vendor-specific implementations of middleware technologies such as Borland's VisiBroker, IBM's WebSphere, and Microsoft's .NET.

Platform independence is a quality of a model that is measured in degrees [60]. The platform specific viewpoint provides a view of a system in which platform specific details are integrated with the elements in a PIM. This view of a system is described by a platform specific model (PSM). Separation of platform specific and platform independent details is considered the key to providing effective support for migrating an application from one implementation platform to another.

The pillars of MDA are the Meta Object Facility (MOF), a language for defining the abstract syntax of modeling languages [44], the UML [59], and the Query, View, Transformation standard (QVT), a standard for specifying and implementing model transformations (e.g., PIM to PSM transformations) [46].

### 4.2. Software Factories

Information about the Microsoft Software Factory initiative became widely available when a book on the topic was published in 2004 [28]. The initiative focuses on the development of MDE technologies that leverage domain-specific knowledge to automate software modeling tasks.

Software Factories tackle the complexity of bridging the gap by providing developers with a framework for producing development environments consisting of domain-specific tools that help automate the transformation of abstract models to implementations. Each development environment is defined as a graph of viewpoints, where each viewpoint describes systems in the application domain from the perspective of some aspect of the development lifecycle (e.g., from a requirements capture or a database design perspective). Reusable forms of development experience (e.g., patterns, templates, guidelines, transformations) are associated with each viewpoint, and thus accessible in the context of that viewpoint. This reduces the need to search for applicable forms of reusable experience, and enables context-based validation, and guidance delivery and enactment [28].

The relationships between viewpoints define semantic links between elements in the viewpoints. For example, the relationships can be used to relate work carried out in one development phase with work performed in another phase, or to relate elements used to describe an aspect of the system with elements used to describe a different aspect. In summary, the Software Factory initiative is concerned with developing technologies that can be used to create development environments for a family of applications.

There are three key elements in a realization of the Software Factory vision:

- **Software factory schema:** This schema is a graph of viewpoints defined using Software Factory technologies. It describes a product line architecture in terms of DSMLs to be used, and the mechanisms to be used to transform models to other models or to implementation artifacts.

- **Software factory template:** A factory template provides the reusable artifacts, guidelines, samples, and custom tools needed to build members of the product family.

- **An extensible development environment:** To realize the software factory vision, a framework that can

be configured using the factory schema and template to produce a development environment for a family of products is needed. The Microsoft Visual Studio Team System has some elements of this framework and there is ongoing work on extending its capabilities.

### 4.3. Other MDE Approaches

Other notable work in the domain-specific modeling vein is the work of Xactium on providing support for engineering domain-specific languages (see http://www.xactium.com), and the work at Vanderbilt University on the Generic Modeling Environment (GME) (see http://www.isis.vanderbilt.edu/projects/gme/). Both approaches are based on the MOF standard of MDA and provide support for building MOF-based definitions of domain-specific modeling languages. The MOF is used to build models, referred to as *metamodels*, that define the abstract syntax of modeling languages. While these approaches utilize MDA standards they do not necessarily restrict their modeling viewpoints to the CIM, PIM and PSM. The Xactium approach, in particular, is based on an adaptive tool framework that uses reflection to adapt a development environment as its underlying modeling language changes: If extensions are made to the modeling language the environment is made aware of it through reflection and can thus adapt.

A major MDE initiative from academia is *Model Integrated Computing* (MIC) [57]. The MIC initially started out with a focus on developing support for model driven development of distributed embedded real-time systems. There is now work taking place within the OMG to align the MIC and MDA initiatives (e.g., see http://mic.omg.org/).

## 5. Modeling Language Challenges

The following are two major challenges that architects of MDE modeling languages face:

- *The abstraction challenge*: How can one provide support for creating and manipulating problem-level abstractions as first-class modeling elements in a language?

- *The formality challenge*: What aspects of a modeling language's semantics need to be formalized in order to support formal manipulation, and how should the aspects be formalized?

Two schools of thought on how to tackle the abstraction challenge have emerged in the MDE community:

- *The Extensible General-Purpose Modeling Language School:* The abstraction challenge is tackled by providing a base general-purpose language with facilities to extend it with domain-specific abstractions (i.e., abstractions that are specific to a problem domain).

- *The Domain Specific Modeling Language School:* The challenge is tackled by defining domain specific languages using meta-metamodeling mechanisms such as the OMG's MOF. The focus of work in this area is on providing tool support for engineering modeling languages. The products of Xactium, MetaCase, and Microsoft provide examples of current attempts at producing such tools.

It is important to note that the research ideas, techniques and technologies from these two schools are not mutually exclusive. Extensible modeling languages and meta-metamodeling technologies can both play vital roles in an MDE environment. We envisage that research in both schools will provide valuable insights and research results that will lead to a convergence of ideas.

### 5.1. Learning from the UML Experience: Managing Language Complexity

> "It is easier to perceive error than to find truth, for the former lies on the surface and is easily seen, while the latter lies in the depth, where few are willing to search for it." – Johann Wolfgang von Goethe

An extensible, general-purpose modeling language should provide, at least, (1) abstractions above those available at the code level that support a wide variety of concepts in known problem domains, and (2) language extension mechanisms that allow users to extend or specialize the language to provide suitable domain-specific abstractions for new application domains.

The Extensible Modeling Language school is exemplified by work on the UML. There are significant benefits to having a standardized extensible general-purpose modeling language such as the UML. For example, such a language facilitates communication across multiple application domains and makes it possible to train modelers that can work in multiple domains.

The UML is also one of the most widely critiqued modeling languages (e.g., see [22, 31]). Despite its problems, there is no denying that the UML standardization effort is playing a vital role as a public source of insights into problems associated with developing practical software modeling languages.

A major challenge that is faced by developers of extensible general-purpose modeling language is identifying a small base set of modeling concepts that can be used to express a wide range of problem abstractions. The UML standardization process illustrates the difficulty of converging on a small core set of extensible concepts. One of the problems is that there is currently very little analyzable modeling experience that can be used to distill a small core of extensible modeling concepts. One way to address this problem is to set up facilities for collecting, analyzing and sharing modeling experience, particularly from industry. There are a number of initiatives that seek to develop and maintain a repository of modeling experience, for example PlanetMDE (see http://planetmde.org/) and REMODD (see http://lists.cse.msu.edu/cgi-bin/mailman/listinfo/remodd). Collecting relevant experience from industry will be extremely challenging. Assuring that Intellectual Property rights will not be violated and overcoming the reluctance of organizations to share artifacts for fear that analysis will reveal embarrassing problems are some of the challenging problems that these initiatives must address.

The complexity of languages such as the UML is reflected in their metamodels. Complex metamodels are problematic for developers who need to understand and use them. These include developers of MDE tools and transformations. The complexity of metamodels for standard languages such as the UML also presents challenges to the groups charged with evolving the standards [22]. An evolution process in which changes to a metamodel are made and evaluated manually is tedious and error prone. Manual techniques make it difficult to (1) establish that changes are made consistently across the metamodel, (2) determine the impact changes have on other model elements, and (3) determine that the modified metamodel is sound and complete. It is important that metamodels be shown to be sound and complete. Conformance mechanisms can then be developed and used by tool vendors to check that their interpretations of rules in the metamodel are accurate.

Tools can play a significant role in reducing the accidental complexities associated with understanding and using large metamodels. For example, a tool that extracts metamodel views of UML diagram types consisting only of the concepts and relationships that appear in the diagrams can help one understand the relationships between visible elements of a UML diagram. A more flexible and useful approach is to provide tools that allow developers to query the metamodel and to extract specified views from the metamodel. Query/Extraction tools should be capable of extracting simple derived relationships between concepts and more complex views that consist of derived relationships among many concepts. Metamodel users can use such tools to better understand the UML metamodel, and to obtain metamodel views that can be used in the specification of patterns and transformations. Users that need to extend or evolve the UML metamodel can also use such tools to help determine the impact of changes (e.g., a query that returns a view consisting of all classes directly or indirectly related to a concept to be changed in a metamodel can provide useful information) and to check that changes are consistently made across the metamodel. The development of such tools is not beyond currently available technologies. Current UML model development tools have some support for manipulating the UML metamodel that can be extended with query and extraction capabilities that are accessible by users.

Another useful tool that can ease the task of using the UML metamodel is one that takes a UML model and produces a metamodel view that describes its structure. Such a tool can be used to support compliance checking of models.

### 5.2. Learning from the UML Experience: Extending Modeling Languages

The UML experience provides some evidence that defining extension mechanisms that extend more than just the syntax of a language is particularly challenging. UML 2.0 supports two forms of extensions: Associating particular semantics to specified semantic variation points, and using profiles to define UML variants.

A semantic variation point is a semantic aspect of a model element that the UML allows a user to define. For example, the manner in which received events are handled by a state machine before processing is a semantic variation point for state machines: A modeler can decide to use a strict queue mechanism, or another suitable input handling mechanism. A problem with UML semantic variation points is that modelers are responsible for defining and communicating the semantics to model readers and tools that manipulate the models (e.g., code generators). UML 2.0 does not provide default semantics or a list of possible variations, nor does it formally constrain the semantics that can be plugged into variation points. This can lead to the following pitfalls: (1) Users can unwittingly assign a semantics that is inconsistent with the semantics of related concepts; and (2) failure to communicate a particular semantics to model readers and to tools that analyze models can lead to misinterpretation and improper analysis of the models. The challenge here is to develop support for defining, constraining and checking the semantics plugged into semantic variation points.

Profiles are the primary mechanism for defining domain-specific UML variants. A UML profile describes how UML model elements are extended to support usage in a particular domain. For example, a profile can be used to define a variant of the UML that is suited for modeling J2EE software systems. UML model elements are extended using stereotypes and tagged values that define additional properties that are to be associated with the elements. The extension of a model element introduced by a stereotype must not contradict the properties associated with the model element. A profile is a lightweight extension mechanism and thus cannot be used to add new model elements or delete existing model elements. New relationships between UML model elements can be defined in a profile though.

The OMG currently manages many profiles including the Profile for Schedulability, Performance and Time and a system modeling profile called SysML. Unfortunately, the UML 2.0 profile mechanism does not provide a means for precisely defining semantics associated with extensions. For this reason, profiles cannot be used in their current form to develop domain-specific UML variants that support the formal model manipulations required in an MDE environment. The XMF-Mosaic tool developed by Xactium takes a promising approach that is based on the use of meta-profiles and a reflective UML modeling environment that is able to adapt to extensions made to the UML.

### 5.3. Domain Specific Modeling Environments

A domain specific language consists of constructs that capture phenomena in the domain it describes. Domain specific languages (DSL) cover a wide range of forms [15]. A DSL may be used for communication between software components (e.g., XML-dialects), or it may be embedded in a wizard that iteratively asks a user for configuration information.

DSLs can help bridge the problem-implementation gap, but their use raises new challenges:

- Enhanced tooling challenge: Each DSL needs its own set of tools (editor, checker, analyzers, code generators). These tools will need to evolve as the domain evolves. Building and evolving these tools using manual techniques can be expensive. A major challenge for DSL researchers is developing the foundation needed to produce efficient meta-toolsets for DSL development.

- The DSL-Babel challenge: The use of many DSLs can lead to significant interoperability, language-version and language-migration problems. This problem poses its own challenges with respect to training and communication across different domains. DSLs will evolve and will be versioned and so must the applications that are implemented using the DSLs. Furthermore, different parts of the same system may be described using different DSLs and thus there must be a means to relate concepts across DSLs and a means to ensure consistency of concept representations across the languages. Sound integration of DSLs will probably be as hard to achieve as the integration of various types of diagrams in a UML model.

The developers responsible for creating and evolving DSL tools will need to have intimate knowledge of the domain and thus must closely interact with application developers. Furthermore, the quality of the DSL tools should be a primary concern, and quality assurance programs for the DSL tooling sector of an organization should be integrated with the quality assurance programs of the application development sectors. These are significant process and management challenges.

In addition to these challenges many of the challenges associated with developing standardized modeling languages apply to DSLs.

### 5.4. Developing Formal Modeling Languages

Formal methods tend to restrict their modeling viewpoints in order to provide powerful analysis, transformation and generation techniques. A challenge is to integrate formal techniques with MDE approaches that utilize modeling languages with a rich variety of viewpoints. A common approach is to translate a modeling view (e.g. a UML class model) to a form that can be analyzed using a particular formal technique (e.g., see [42]). For example, there are a number of approaches to transforming UML design views to representations that can be analyzed by model checking tools. Challenges here are, to (a) ensure that the translation is semantically correct, and (b) hide the complexities of the target formal language and tools from the modeler. Meeting the latter challenge involves automatically translating the analysis results to a form that utilizes concepts in the source modeling language.

Another approach would be to integrate the analysis/generation algorithms with the existing modeling language. This is more expensive, but would greatly enhance the applicability of an analysis tool to an existing modeling language.

In the formal methods community the focus is less on developing new formal languages and more on tuning existing notations and techniques. MDE languages provide a good context for performing such tuning.

### 5.5. Analyzing Models

If models are the primary artifacts of development then one has to be concerned with how their quality is evaluated. Good modeling methods should provide modelers with criteria and guidelines for developing quality models. These guidelines can be expressed in the form of patterns (e.g., Craig Larman's GRASP patterns), proven rules of thumb (e.g., "minimize coupling, maximize cohesion", "keep inheritance depth shallow"), and exemplar models. The reality is that modelers ultimately rely on feedback from experts to determine "goodness" of their models. For example, in clasrooms the instructors play the role of expert modelers and students are apprentices. From the student perspective, the grade awarded to a model reflects its "goodness". The state of the practice in assessing model quality provides evidence that modeling is still in the craftsmanship phase.

Research on rigorous assessment of model quality has given us a glimpse of how we can better evaluate model quality. A number of researchers are working on developing rigorous static analysis techniques that are based on well-defined models of behaviors. For example, there is considerable work on model-checking of modeled behavior (e.g., see [39]).

Another promising area of research is systematic model testing. Systematic code testing involves executing programs on a select set of test inputs that satisfy some test criteria. These ideas can be extended to the modeling phases when executable forms of models are used. Model testing is concerned with providing modelers with the ability to animate or execute the models they have created in order to explore the behavior they have modeled.

The notion of model testing is not new. For example, SDL (Specification and Description Language) tools provide facilities for exercising the state-machine based models using an input set of test events. Work on executable variants of the UML also aims to provide modelers with feedback on the adequacy of their models. More recently a small, but growing, number of researchers have begun looking at developing systematic model testing techniques. This is an important area of research and helps pave the way for more effective use of models during software development.

There may be lessons from the systematic code testing community that can be applied, but the peculiarities of modeling languages may require the development of innovative approaches. In particular, innovative work on defining effective test criteria that are based on coverage of model elements and on the generation of model-level test cases that provide desired levels of coverage is needed.

The ability to animate models can help one better understand modeled behavior. Novices and experienced developers will both benefit from the visualization of modeled behavior provided by model animators. Model animation can give quick visual feedback to novice modelers and can thus help them identify improper use of modeling constructs. Experienced modelers can use model animation to understand designs created by other developers better and faster.

It may also be useful to look at how other engineering disciplines determine the quality of their models. Engineers in other disciplines typically explore answers to the following questions when determining the adequacy of their models: Is the model a good predictor of how the physical artifact will behave? What are the assumptions underlying the model and what impact will they have on actual behavior? The answer to the first question is often based on evidence gathered from past applications of the model. Evidence of model fidelity is built up by comparing the actual behavior of systems built using the models with the behavior predicted by the models. Each time engineers build a system the experience gained either reinforces their confidence in the predictive power of the models used or the experience is used to improve the predictive power of models. Answers to the second question allow engineers to identify the limitations of analysis carried out using the models and develop plans for identifying and addressing problems that arise when the assumptions are violated.

Are similar questions applicable to software models? There are important differences between physical and software artifacts that one needs to take into consideration when applying modeling practices in other engineering disciplines to software, but there may be some experience that can be beneficially applied to software modeling.

## 6. Supporting Separation of Design Concerns

Developers of complex software face the challenge of balancing multiple interdependent, and sometimes conflicting, concerns in their designs. Balancing pervasive dependability concerns (e.g., security and fault tolerance concerns) is particularly challenging: The manner in which one concern is addressed can limit how other concerns are addressed, and interactions among software features[2] that address the concerns can give rise to undesirable emergent behavior. Failure to identify and address faults arising from interacting dependability features during design can lead to costly system failures. For example, the first launch of the space shuttle Columbia was delayed because "(b)ackup flight software failed to synchronize with primary avionics software system" (see http://science.ksc.nasa.gov/shuttle/missions/sts-1/mission-sts-1.html). In this case, features that were built in to address fault-tolerance concerns did not interact as required with the primary functional features. Design modeling techniques should allow developers to separate these features so that their interactions can be analyzed to identify faulty interactions and to better understand how emergent behavior arises.

Modeling frameworks such as the MDA advocate modeling systems using a fixed set of viewpoints (e.g., the CIM, PIM, and PSM MDA views). Rich modeling languages such as the UML provide good support for modeling systems from a fixed set of viewpoints. Concepts used in a UML viewpoint are often dependent on concepts used in other viewpoints. For example, participants in a UML interaction diagram must have their classifiers (e.g., classes) defined in a static structural model. Such dependencies are specified in the language metamodel and thus the metamodel should be the basis for determining consistency of information across system views. Unfortunately, the size and complexity of the UML 2.0 metamodel makes it extremely difficult for tool developers and researchers to fully identify the dependencies among concepts, and to determine whether the metamodel captures all required dependencies. In the previous section we discussed the need for tools that query and navigate metamodels of large languages such as the UML. These tools will also make it easier to develop mechanisms that check the consistency of information across views.

---

[2] In this paper, a feature is a logical unit of behavior.

The fixed set of viewpoints provided by current modeling languages and frameworks such as the MDA and UML are useful, but more is needed to tackle the complexity of developing software that address pervasive interdependent concerns. The need for better separation of concerns mechanisms arises from the need to analyze and evolve interacting pervasive features that address critical dependability concerns. A decision to modularize a design based on a core set of functional concerns can result in the spreading and tangling of dependability features in a design. The tangling of the features in a design complicates activities that require understanding, analyzing, evolving or replacing the crosscutting features. Furthermore, trade-off analysis tasks that require the development and evaluation of alternative forms of features are difficult to carry out when the features are tangled and spread across a design. These crosscutting features complicate the task of balancing dependability concerns in a design through experimentation with alternative solutions.

Modeling languages that provide support for creating and using concern-specific viewpoints can help alleviate the problems associated with crosscutting features. Developers can use a concern-specific viewpoint to create a design view that describes how the concern is addressed in a design. For example, developers can use an access control security viewpoint to describe access control features at various levels of abstraction.

A concern-specific viewpoint should, at least, consist of (1) modeling elements representing concern-specific concepts at various levels of abstractions, and (2) guidelines for creating views using the modeling elements. To facilitate their use, the elements can be organized as a system of patterns (e.g., access control patterns) or they can be used to define a domain-specific language (DSL) for the concern space. For example, a DSL for specifying security policies can be used by developers to create views that describe application-specific security policies. Supporting the DSL approach requires addressing the DSL challenges discussed in Section 5. Furthermore, the need to integrate views to obtain a holistic view of a design requires the development of solutions to the difficult problem of integrating views expressed in different DSLs. One way to integrate these views is to define a metamodel that describes the relationships among concepts defined in the different viewpoints. An interesting research direction in this respect concerns the use of ontolgies to develop such metamodels. An ontology describes relationships among concepts in a domain of discourse. One can view a metamodel as an ontology and thus we should be able to leverage related work on integrating ontologies in work on integrating views described using different DSLs.

Another approach to supporting the definition and use of concept-specific viewpoints is based on the use of *aspect-oriented modeling* (AOM) techniques. These approaches describe views using general-purpose modeling languages and provide mechanisms for integrating the views. In this section we discuss the AOM approach in more detail and present some of the major challenges that must be met to realize its research goals.

### 6.1. Separating Concerns using Aspect Oriented Modeling Techniques

Work on separating crosscutting functionality at the programming level has led to the development of aspect-oriented programming (AOP) languages such as Aspect-J [35]. Work on modeling techniques that utilize aspect concepts can be roughly partitioned into two categories: Those that provide techniques for modeling aspect-oriented programming (AOP) concepts [36], and those that provide requirements and design modeling techniques that tackle the problem of isolating features in modeling views and analyzing interactions across the views. Work in the first category focuses on modeling AOP concepts such as join points and advise using either lightweight or heavyweight extensions of modeling languages such as the UML (e.g., see [13, 37, 41, 56, 55]). These approaches lift AOP concepts to the design modeling level and thus ease the task of transforming design models to AOP programs. On the other hand, these approaches utilize concepts that are tightly coupled with program-level abstractions supported by current AOP languages.

Approaches in the second category (e.g., see [1, 5, 12, 23, 47]) focus more on providing support for separating concerns at higher levels of abstractions. We refer to approaches in this category as AOM approaches. The Theme approach and the AOM approach developed by the Colorado State University (CSU) AOM group exemplify work on AOM. In these approaches, aspects are views that describe how a concern is addressed in a design. These views are expressed in the UML and thus consist of one or more models created using the UML viewpoints.

The model elements in aspect and primary models provide partial views of design concepts. For example, a class representing a replicated resource in a class model of an aspect describing a replication fault tolerance feature will consist only of attributes and operations needed to describe replication and fault tolerance behavior, while the class representation of the resource in the primary model will include attributes and operations describing the core functionality of the resource. A more holistic view of the resource concepts is obtained by merging its partial representations in the aspect and primary models.

AOM approaches provide support for composing aspect and primary models to obtain an integrated design view that can be used to (1) better understand the interactions across the composed design views, (2) analyze interactions to identify conflicts and undesirable emergent behaviors, and (3) generate non-aspect oriented implementations in model-driven engineering (MDE) environments.

Current composition techniques are based on rules for syntactically matching elements across aspect and primary models, which makes it possible to fully automate the composition. The matching rules use syntactic properties (e.g., model element name) to determine whether two model elements represent the same concept or not. For example, a matching rule stating that classes with the same name represent the same concept can be used to merge classes with the same name but different attributes and operations. The composed class will contain the union of the attributes and operations found in the classes that are merged. This reliance on an assumed correspondence between syntactic properties and the concepts represented by model elements can lead to conflict and other problems when it does not exist. There is a need to take into consideration semantic properties, expressed as constraints or as specifications of behavior, when matching model elements.

Consideration of semantic properties is also needed to support verifiable composition. Composition is carried out in a verifiable manner when it can be established that the model it produces has specified properties. A composition tool should be able to detect when it has failed to establish or preserve a specified property and report this to the modeler. Such checks cannot be completely automated, but it may be possible to provide automated support for detecting particular types of semantic conflicts and other interaction problems.

Another major challenge faced by AOM researchers is concerned with integrating AOM techniques into the software development process. Evolution and transformation of models consisting of multiple interrelated views becomes more complex if the necessary infrastructure for managing the views is not present. The challenges associated with developing such an infrastructure are discussed in Section 7.2.

## 6.2. Related Research on Requirements Views and Feature Interactions

Work on requirements and architecture viewpoints [52], and on the feature interaction problem [11] can provide valuable insights that can be used to understand the challenges of separating design concerns and of analyzing interactions. The terms views and viewpoints tend to be associated with work on requirements analysis, but they can also be applied to designs. A design concern such as access control can be considered to be a design viewpoint. Such a viewpoint can provide concepts, patterns or a language that can be used to create design views that describe features addressing the design concern.

Work on feature interactions has tended to focus on features that provide services of value to software users. For example, in the telecommunication industry a feature is a telecommunication service such as call-forwarding, and research on the feature interaction problem in this domain is concerned with identifying undesirable emergent behaviors that arise when these services interact. There is a growing realization that the feature interaction problem can appear in many forms in software engineering. The problem of analyzing interactions among features that address dependability and other design concerns is another variant of the feature interaction problem. One can also consider work on analyzing interactions across views as a form of the feature interaction problem.

Collaborative research involving members from the AOM, the formal methods, the feature interaction and the viewpoint analysis communities is needed to address the challenging problems associated with separating concerns and integrating overlapping views.

## 7. Manipulating Models

Current MDE technologies provide basic support for storing and manipulating models. Environments typically consist of model editors which can detect some syntactic inconsistencies, basic support for team development of models, and limited support for transforming models. Much more is needed if MDE is to succeed. For example, there is a need for rigorous transformation modeling and analysis techniques, and for richer repository-based infrastructures that can support a variety of model manipulations, can maintain traceability relationships among a wide range of models, and can better support team based development of models. In this section we discuss some of the major MDE challenges related to providing support for manipulating and managing models. The section also includes a discussion on the use of models to support runtime activities.

## 7.1. Model Transformation Challenges

A (binary) model transformation defines a relationship between two sets of models. If one set of models is designated as a source set and the other as a target set then a mechanism that implements such a transformation will take the source set of models and produce the target set of models. These are called *operational transformations* in this paper. Model refinement, abstraction, and refactoring are well-known forms of operational transformations. Other forms that will become more widely used as MDE matures are (1) model composition in which the source models representing different views are used to produce a model that integrates the views, (2) model decomposition in which a single model is used to produce multiple target models, and (3) model translation in which a source set of models are transformed to target models expressed in a different language. In particular, model translations are used to transform model created for one purpose to models that are better suited for other purposes. Examples of translations can be found in work on transforming UML models to artifacts that can be formally analyzed using existing analysis tools. This include work on transforming UML to formal specification languages such as Z and Alloy, to performance models, and to state machine representations that can be analyzed by existing model checkers.

Transformations can also be used to maintain relationships among sets of models: Changes in the models in one set trigger changes in the other sets of models in order to maintain specified relationships. These *synchronization transformations* are used to implement change synchronization mechanisms in which changes to a model (e.g., a detailed UML design model) trigger corresponding changes in related artifacts (e.g., code generated from the UML design model).

Research on model transformations is still in its infancy and there is very little experience that can be used to determine the worth of current approaches. The OMG's Query, View, Transformation (QVT) standard defines three types of transformation languages: two declarative languages that describe relations at different levels of abstraction, and an operational transformation language that describes transformations in an imperative manner. In addition to the QVT, there are a number of other proposals for transformation languages. An informative survey of transformation language features can be found in the paper by Czarnecki [16].

More research is needed on analyzing model transformations. The complex structure of models poses special challenges in this respect. As mentioned previously, a model is a collection of interrelated views. The following are some of the difficult research questions that arise from the multi-view nature of models:

- How does one maintain consistency across views as they are transformed? Synchronization transformation technologies may be used here to "ripple" the results of transformations to related views.

- How can transformations be tested? The complex structure of the models may stretch the limits of current formal static analysis and testing techniques. For testing techniques, the complex structures make definition of oracles and effective coverage criteria particularly challenging.

A particular challenge faced by developers of model-to-code transformations is integrating generated code with handcrafted or legacy code. Current code generation tools assume that generated code is stand-alone and provide very little support for integration of foreign code. Integrating foreign and generated code is easier if they are architecturally compatible. Unfortunately, current code generators do not make explicit the architectural choices that are made by the generators when they produce code and provide very limited support for affecting those choices. This makes it difficult to guarantee that a code generator will produce code that is architecturally compatible with foreign code. The result is that some refactoring of the generated and foreign code may be needed, or a separate "glue" interface needs to be developed. It may be possible to generate the needed refactoring steps or the "glue" code given appropriate information about the foreign and generated code. Research on techniques for generating these artifacts will have to determine the needed information.

## 7.2. Model Management Challenges

In a project, many models at varying levels of abstractions are created, evolved, analyzed and transformed. Manually tracking the variety of relationships among the models (e.g., versioning, refinement, realization and dependency relationships) adds significant accidental complexity to the MDE development process. Current modeling tools do not provide the support needed to effectively manage these relationships.

An MDE repository must have the capability to store models produced by a variety of development tools, and must be open and extensible in order to support a close approximation of the MDE vision. The repository should (1) allow tools from a variety of vendors to manipulate the models, (2) monitor and audit the model manipulations, and (3) automatically extract information from the audits and use it to establish, update or remove relationships among models. Developing such a repository requires addressing difficult technical problems. Problems associated with maintaining artifact traceability relationships are notoriously challenging and two decades of research on these problems have not produced convenient solutions.

Research in the area of *Mega-Modeling*, in which models are the units of manipulation, targets the problems associated with managing and maniuplating models [9]. Metamodels play a critical role in mega-modeling: Mechanisms that manipulate models work at the metamodel level and information about artifacts stored in a repository can be captured in their metamodels.

Metamodels need to define more than just the abstract syntax of a language if they are to support model management tools. For example, the UML 2.0 metamodel is a class model in which the classes have get, set and helper functions that are used only to specify the abstract syntax and well-formedness rules. Metamodels should be able to use the full power of modeling languages to define both syntactic and semantic aspects of languages. For example, one can define semantics that determine how models in a language are to be transformed by including supporting operations in the metamodel. Furthermore, the metamodel does not have to consist only of a class model. One can use behavioral models (e.g., activity and sequence models) to describe the manipulations that can be carried out on models.

Tools that manipulate models can be associated with metamodels that describe how manipulations are implemented. This information can be used by MDE model management environments to extract information needed to maintain relationships among models that are manipulated by the tools. The KerMeta tool is an example of a new generation of MDE tools that allows one to extend metamodels with operations defining model manipulations [45, 61].

### 7.3. Supporting the use of Models During Runtime: Opportunities and Challenges

Examples of how the runtime models can be used by different system stakeholders are given below:

- System users can use runtime models to observe the runtime behavior when trying to understand a behavioral phenomenon (e.g., understanding the conditions under which transaction bottlenecks occur in a system), and to monitor specific aspects of the runtime environment (e.g., monitoring patterns of access to highly-sensitive information).

- Adaptation agents can use runtime models to detect the need for adaptation and to effect the adaptations. Effecting an adaption involves making changes to models of the parts to be adapted and submitting the changes to an adaptation mechanism that can interpret and perform the needed adaptations. Here it is assumed that the adaptations to be performed are pre-determined.

- In more advanced systems, change agents (maintainers or software agents) can use the runtime models to correct design errors or to introduce new features to a running system. Unlike adaptations, these changes are not pre-determined and thus the mechanisms used to effect the changes can be expected to be more complex. These more complex mechanisms will be able to support pre-determined adaptations as well as unanticipated modifications.

Research on providing support for creating and using runtime models is in its infancy. At the MODELS workshop on runtime models, Gordon Blair identified the following research questions: What forms should runtime models take? How can the fidelity of the models be maintained? What role should the models play in validation of the runtime behavior? These questions are good starting points for research in the area.

Providing support for changing behavior during runtime is particularly challenging. If models are to be used to effect changes to running software, research needs to focus on how the changes can be effected in a controlled manner. Allowing developers to change runtime behavior in an ad-hoc manner is obviously dangerous. A model-based runtime change interface will have to constrain how changes are effected and provide the means to check the impact of change before applying it to the running system.

## 8. Conclusions

In this paper we suggest that MDE research focus on providing technologies that address the recurring problem of bridging the problem-implementation gap. We also encourage research on the use of runtime models. The problems and challenges outlined in this paper are difficult to overcome, and it may seem that MDE techniques are more likely to contribute to the complexity of software development rather than manage inherent software complexity. It is our view that software engineering is inherently a modeling activity, and that the complexity of software will overwhelm our ability to effectively maintain mental models of a system. By making the models explicit and by using tools to manipulate, analyze and manage the models and their relationships, we are relieving significant cognitive burden and reducing the accidental complexities associated with maintaining mentally held models.

The web of models maintained in an MDE environment should be a reflection of inherent software complexity. Currently, creating software models is an art and thus models of faulty or convoluted solutions, and messy descriptions of relatively simple solutions can be expected. These modeling problems will give rise to accidental complexities.

There will always be accidental complexities associated with learning and using modeling languages and MDE tools to develop complex software. MDE technologists should leverage accumulated experience and insights gained from failed and successful applications of previous MDE technologies to develop new technologies that reduce the accidental complexities of the past.

To conclude the paper we present a vision of an MDE environment that, if realized, can conceivably result in order-of-magnitude improvement in software development productivity and quality. The vision is intentionally ambitious and may not be attainable in its entirety. Progressively closer approximations of the vision will have increasingly significant effects on the effort required to develop complex software. In this sense, the vision can act as a point of reference against which MDE research progress can be informally assessed.

In the MDE vision, domain architects will be able to produce domain specific application development environments (DSAEs) using what we will refer to as MDE technology frameworks. Software developers will use DSAEs to produce and evolve members of an application family. A DSAE consists of tools to create, evolve, analyze, and transform models to forms from which implementation, deployment and runtime artifacts can be generated. Models are stored in a repository that tracks relationships across modeled concepts and maintains metadata on the manipulations that are performed on models.

Some of the other features that we envisage will be present in a DSAE are (1) mechanisms supporting for round-trip engineering, (2) mechanisms for synchronizing models at different levels of abstraction when changes are made at any level, and (3) mechanisms for integrating generated software with legacy software. Developers should also be able to use mechanisms in the DSAE to incorporate software features supporting the generation and use of runtime models.

Realizing the MDE vision of software engineering is a wicked problem and thus MDE environments that lead to order-of-magnitude improvements in software productivity and quality are not likely to appear in the near to medium-term - barring new insights that could lead to significant improvement in the rate at which pertinent technologies are developed. The current limitations of MDE technologies reflect inadequacies in our understanding of the software modeling phenomenon. The development and application of progressively better MDE technologies will help deepen our understanding and move us closer to better approximations of the MDE vision.

**Acknowledgments:** Robert France's work on this paper was supported by a Lancaster University project VERA: Verifiable Aspect Models for Middleware Product Families, funded by the UK Engineering and Physical Sciences Research Council (EPSRC) Grant EP/E005276/1. This work is also partly undertaken within the MODELPLEX project funded by the EU under the IST Programme. The authors thank the editors and following persons for their valuable feedback on drafts of the paper: Nelly Bencomo, Gordon Blair, Betty Cheng, Tony Clarke, Steve Cook, Andy Evans, Awais Rashid, Bran Selic, Richard Taylor, Laurie Tratt.